\documentclass[a4paper]{jpconf}
\usepackage{graphicx}

\usepackage{amsmath}
\makeatletter
\newcommand*\rel@kern[1]{\kern#1\dimexpr\macc@kerna}
\newcommand*\widebar[1]{%
	\begingroup
	\def\mathaccent##1##2{%
		\rel@kern{0.8}%
		\overline{\rel@kern{-0.8}\macc@nucleus\rel@kern{0.2}}%
		\rel@kern{-0.2}%
	}%
	\macc@depth\@ne
	\let\math@bgroup\@empty \let\math@egroup\macc@set@skewchar
	\mathsurround\z@ \frozen@everymath{\mathgroup\macc@group\relax}%
	\macc@set@skewchar\relax
	\let\mathaccentV\macc@nested@a
	\macc@nested@a\relax111{#1}%
	\endgroup
}
\makeatother

\def\){\right)}
\def\({\left(}

\def\non{\nonumber}
\def\ba{\begin{align}}
\def\ea{\end{align}}
\def\br{\begin{eqnarray}}
\def\er{\end{eqnarray}}

\def\d{\delta}

\def\s{\sigma}
\def\bp{\bar{\psi}}

\def\ph{\phi}
\def\p{\psi}
\def\pp{\partial}

\def\vph{\varphi}
\def\bchi{\bar{\chi}}
\def\cn{{\cal N}}
\def\cl{{\cal L}}
\def\ep{\epsilon}
\def\bep{\bar{\epsilon}}

\begin{document}
\title{Fusing defect for the ${\cal N} = 2$ super sinh-Gordon model}

\author{N I Spano, A R Aguirre, J F Gomes and A H Zimerman}

\address{Rua Dr. Bento Teobaldo Ferraz 271, Block 2, 01140-070, S\~ao Paulo, Brazil}

\ead{natyspano@ift.unesp.br}

\begin{abstract}
In this paper we derive  the type-II integrable defect for the ${\cal N} = 2$ supersymmetric sinh-Gordon (sshG) model by using the fusing procedure. In particular, we show explicitly the conservation of the modified energy, momentum and supercharges.

\end{abstract}

%%%%%%%%%%%%%%%%%%%%%%%%%%%%%%%%%%%%%%%%%%%%%%%%

\section{Introduction}
An interesting topic in the study of integrable systems is the analysis of their integrability properties in the presence of  impurities or defects. Accordingly, defects are introduced in two-dimensional integrable field theories as internal boundary conditions located at a fixed point in the $x$-axis, which connect two different field theories of both sides of it. In particular, after the introduction of the defect, the spatial translation invariance is broken since some constraints are imposed to be satisfied at a particular space point, and hence it would be expected a violation of momentum conservation. However, it was verified in \cite{1}-\cite{6}, that in order to preserve the integrability, the fields of the theory  must satisfy a kind of B\"acklund transformation frozen at the defect point.

This kind of  integrable defects can be classified into two classes: \emph{type-I}, if the fields on both sides only interact with each other at the defect point, and \emph{type-II} if they interact through  additional degrees of freedom present only at the defect point \cite{4}. The type-II formulation proved to be suitable not only for describing defects within the Tzitz\'eica-Bullough-Dodd ($a_2^{(2)}$-Toda) model, which had been excluded from the type-I setting, but it also provided additional types of defects for the sine-Gordon (sG) and others affine Toda field theories (ATFT) \cite{Corr5}. Interestingly, for the sG model \cite{4, Corr10}, and in general for $a_r^{(1)}$-ATFT \cite{CR} and the ${\cn}=1$ sshG models \cite{Nathaly}, the type-II defects can be regarded as  fused pairs of type-I defects  previously placed at different points in space. However, the type-II defects can  be allowed in models that cannot support type-I defects, as it was shown for the $a_2^{(2)}$-Toda model \cite{4}.

On the other hand, the presence of integrable defects  in the ${\cn} =1$ sshG model has been already discussed in \cite{FLZ1, Nathaly14}. However, the kind of defect introduced in those papers can be regarded as a partial type-II defect since only auxiliary fermionic fields appear in the defect Lagrangian, and consequently it reduces to type-I defect for sinh-Gordon model in the bosonic limit. The proper supersymmetric extension of the type-II defect for the $\cn=1$ sshG model was recently proposed in \cite{Nathaly}, by using two methods: the generalization of the super-B\"acklund transformations, and the fusing procedure.

The purpose of this paper is to derive type-II defects for the $\cn = 2$ sshG equation by fusing defects of the kind already known in literature \cite{FLZ2}. The explicit form of the type-II B\"acklund transformations for the $\cn = 2$ sshG model  will be presented. We will also compute its modified conserved energy, momentum and supercharges. Finally, by introducing appropriate field transformations, the ${\bf PT}$ symmetry of  the bulk and the defect theories will be discussed.

\section{${\cal N}=2$ super sinh-Gordon model}
The action for the bulk $\cn=2$ sshG model is given by,
\br
S_{bulk} = \int_{-\infty}^{\infty} dt \int_{-\infty}^{\infty}  dx \ {\cal L}_{bulk},
\er
with the bulk Lagrangian density,
\begin{eqnarray}
{\mathcal L}_{bulk} &=&\frac{1}{2}(\partial_x \phi)^2 - \frac{1}{2}(\partial_t \phi)^2 -\frac{1}{2}(\partial_x \vph)^2 + \frac{1}{2}(\partial_t \vph)^2-i\p(\pp_x-\pp_t)\p+i\bp(\pp_x+\pp_t)\bp\non\\
&&+ i\chi(\pp_x-\pp_t)\chi-i\bchi(\pp_x+\pp_t)\bchi+m^2\big[\cosh(2\ph)-\cosh(2\vph)\big]\non\\
&&+4im(\bp\p+\bchi\chi)\cosh\ph\cosh\vph-4im(\bp\chi+\bchi\p)\sinh\ph\sinh\vph,
\end{eqnarray}
where $\ph$, $\vph$ are bosonic fields, $\p$, $\bp$, $\chi$, $\bchi$ are fermionic fields. Then, the bulk field equations are,
\begin{eqnarray}\label{mov geral 1}
(\pp_{x}^{2}-\pp_{t}^{2})\ph&=& 2m^2\sinh(2\ph)+4im(\bp\p+\bchi\chi)\sinh\ph\cosh\vph\non\\&&-4im(\bp\chi+\bchi\p)\cosh\ph\sinh\vph,\non\\[0.1cm]
(\pp_{x}^{2}-\pp_{t}^{2}) \vph&=& 2m^2\sinh(2\vph)-4im(\bp\p+\bchi\chi)\cosh\ph\sinh\vph\non\\&&+4im(\bp\chi+\bchi\p)\sinh\ph\cosh\vph,\non\\[0.1cm]
(\pp_{x}-\pp_{t})\p &=& -2m\left[\bp\cosh\ph\cosh\vph- \bchi\sinh\ph\sinh\vph  \right], \\[0.1cm]
(\pp_{x}+\pp_{t})\bp &=& -2m\left[\p\cosh\ph\cosh\vph- \chi\sinh\ph\sinh\vph  \right],\non
\\[0.1cm]
(\pp_{x}-\pp_{t})\chi &=& 2m\left[\bchi\cosh\ph\cosh\vph- \bp\sinh\ph\sinh\vph  \right], \non \\[0.1cm]
(\pp_{x}+\pp_{t})\bchi &=& 2m\left[\chi\cosh\ph\cosh\vph- \p\sinh\ph\sinh\vph  \right],\non
\end{eqnarray}
%This equations of motion are invariant under the supersymmetry transformations\\_p
%
The bulk action and the equation of motion have on-shell $\cn=2$ supersymmetry (susy). The susy transformation is given by,
\begin{equation}\label{susy}
	\begin{array}{rcl}
		\d\phi &=& i(\ep_1\psi + \bep_1\bp) - i(\ep_2\chi +\bep_2 \bchi), \\
		\d\vph &=& i(\ep_2\psi + \bep_2\bp) - i(\ep_1\chi +\bep_1 \bchi), \\
		\d\psi &=& (\ep_1 \pp_+\ph + \bep_1 m \sinh\ph \cosh \vph)-(\ep_2 \pp_+\vph + \bep_2 m \sinh\vph \cosh \ph),\\
		\d\chi &=& (\ep_2 \pp_+\ph -\bep_2 m \sinh\ph\cosh\vph)-(\ep_1 \pp_+\vph -\bep_1 m \sinh\vph\cosh\ph),\\
		\d\bp &=& (\bep_2\pp_-\vph +\ep_2 m \sinh\vph\cosh\ph)-(\bep_1\pp_-\ph +\ep_1 m \sinh\ph\cosh\vph),\\
		\d\bchi &=& (\bep_1\pp_-\vph -\ep_1 m \sinh\vph\cosh\ph)-(\bep_2\pp_-\ph -\ep_2 m \sinh\ph\cosh\vph),
	\end{array}
\end{equation}
where $\ep_{k}$ and $\bep_k$, with $k=1,2$, are fermionic parameters, and the light-cone notation $x_\pm =x\pm t$, and \mbox{$\pp_\pm=\frac{1}{2}(\pp_x\pm\pp_t)$} has been used. It can be easily verified that the equations of motions are invariant under these transformations. For simplicity, we will focus on the $\ep_1$-projection of the susy transformation (\ref{susy}), which will be denoted $\d_1$, and then we will compute the associated supercharge $Q_{\ep_1}\equiv Q_1$. 

Under a not-rigid susy transformation, i.e
with parameters $\ep(x, t)$ and $\bep(x, t)$, ${\cal L}_{bulk}$ changes by a total derivative
\br
\d_1{\cal L}_{bulk} &=& \pp_x\Big[ i\ep_1\Big(\psi\(\pp_-\ph+2\pp_+\ph\) +\chi\(\pp_-\vph+2\pp_+\vph\)+m\bp\sinh\ph\cosh\vph -m\bchi\sinh\vph\cosh\ph\Big)\Big] \non\\
&&\!\!\!\!+\pp_t\Big[ i\ep_1\Big(\psi\(\pp_-\ph-2\pp_+\ph\) +\chi\(\pp_-\vph+2\pp_+\vph\)+m\bp\sinh\ph\cosh\vph-m\bchi\sinh\vph\cosh\ph\Big)\Big]\non\\
&&\!\!\!\!+\ep_1\Big[ \pp_t\Big(2i\p\pp_+\ph + 2i\chi\pp_+\vph -2im \bp\sinh\ph \cosh\vph +2im\bchi\sinh\vph\cosh\ph \Big)\non\\
&& \,\,\quad-\pp_x \Big( 2i\p\pp_+\ph + 2i\chi\pp_+\vph +2im \bp\sinh\ph \cosh\vph -2im\bchi\sinh\vph\cosh\ph\Big)\Big],\label{variacao}
\er
if the conservation law inside the last square-bracket in (\ref{variacao}) is hold. Then, the associated bulk supercharges $Q_1$ is given  by an integral of the fermionic density, namely
\br
Q_1 &=& \int_{-\infty}^{\infty} dx \Big[i\p(\pp_x+\pp_t)\ph + i\chi(\pp_x+\pp_t)\vph -2im \bp\sinh\ph \cosh\vph +2im\bchi\sinh\vph\cosh\ph \Big].\quad \mbox{}
\er
The derivation of the remaining supercharges follows the same line of reasoning. Their explicit form is given by the following expressions,
\br
{\widebar Q}_1 &=& \int_{-\infty}^{\infty} dx \Big[i\bp(\pp_x-\pp_t)\ph + i\bchi(\pp_x-\pp_t)\vph -2im \p\sinh\ph \cosh\vph +2im\chi\sinh\vph\cosh\ph \Big],\quad \mbox{}\\
Q_2 &=& \int_{-\infty}^{\infty} dx \Big[i\chi(\pp_x+\pp_t)\ph + i\p(\pp_x+\pp_t)\vph +2im \bchi\sinh\ph \cosh\vph -2im\bp\sinh\vph\cosh\ph \Big],\quad \mbox{}\\
{\widebar Q}_2&=& \int_{-\infty}^{\infty} dx \Big[i\bp(\pp_x-\pp_t)\vph + i\bchi(\pp_x-\pp_t)\ph +2im \chi\sinh\ph \cosh\vph -2im\p\sinh\vph\cosh\ph \Big].\quad \mbox{}
\er
In next section we introduce the Lagrangian description of type-I defects in ${\cal N}=2$ sshG model.

\section{Type-I defect for ${\cal N}=2$ sshG model}
We consider a defect placed in $x=0$ connecting two field theories $\Phi_1$ in the region $x<0$ and $\Phi_2$ in the region $x>0$,
\begin{figure}[h]
\begin{center}
\includegraphics[width=0.5\linewidth]{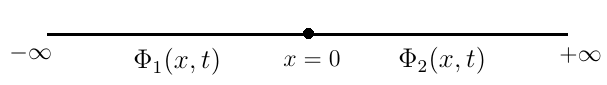}
\end{center} 
\caption{\label{fig1}Defect Representation.}
\end{figure}
First of all, let us consider a Lagrangian density for the region $x<0$ describing the set of fields $\Phi_1(\ph_1,\p_1,\bp_1,\vph_1,\chi_1,\bchi_1)$ and correspondingly $\Phi_2(\ph_2,\p_2,\bp_2,\vph_2,\chi_2,\bchi_2)$ in the region $x>0$, and a defect located at $x=0$, in the following way
\begin{eqnarray}\label{fused}
{\mathcal{L}}=\theta(-x){\mathcal{L}}_{1}+\theta(x){\mathcal{L}}_{2}+\delta(x){\mathcal{L}}_{D},
\end{eqnarray}
where $\mathcal{L}_1$ and $\mathcal{L}_2$ are the bulk Lagrangian densities corresponding to $x < 0$ and $x > 0$ regions, respectively, and the defect Lagrangian density $\mathcal{L}_D$ is given by 
\begin{eqnarray}
{\mathcal L}_{D} &=&\frac{1}{2}(\phi_2\partial_t\phi_1-\phi_1\partial_t\phi_2)-\frac{1}{2}(\vph_2\partial_t\vph_1-\vph_1\partial_t\vph_2)+B_0(\ph_1,\ph_2,\vph_1,\vph_2)\non\\
&&-i(\bp_1\bp_2+\p_1\p_2) +i(\bchi_1\bchi_2+\chi_1\chi_2) + \frac{i}{2}(f\pp_tg+g\pp_tf)\non\\&&+B_1(\ph_1,\ph_2,\vph_1,\vph_2,\p_1,\p_2,\bp_1,\bp_2,\chi_1,\chi_2,\bchi_1,\bchi_2,f,g)
\end{eqnarray}
where $f$, $g$ are fermionic auxiliary fields. For $x=0$, we obtain the following defect conditions,
\begin{equation} \label{mov geral def}
\begin{array}{r c l c r c l}\pp_{x}\phi_{1}-\pp_{t}\phi_{2}&=&-\pp_{\phi_{1}}(B_{0}+B_{1}),&  \mbox{} \qquad & \pp_{x}\vph_{1}-\pp_{t}\vph_{2}&=&\pp_{\vph_{1}}(B_{0}+B_{1}),\\
\pp_{x}\phi_{2}-\pp_{t}\phi_{1}&=&\pp_{\phi_{2}}(B_{0}+B_{1}),&  \mbox{} \qquad &\pp_{x}\vph_{2}-\pp_{t}\vph_{1}&=&-\pp_{\vph_{2}}(B_{0}+B_{1}), \\
i(\p_{1}-\p_{2})&=&-\pp_{\p_{1}}B_{1}=-\pp_{\p_{2}}B_{1},&  \mbox{} \qquad &i(\chi_{1}-\chi_{2})&=&\pp_{\chi_{1}}B_{1}=\pp_{\chi_{2}}B_{1},\\
i(\bp_{1}+\bp_{2})&=&\pp_{\bp_{1}}B_{1}=-\pp_{\bp_{2}}B_{1},&  \mbox{} \qquad &i(\bchi_{1}+\bchi_{2})&=&-\pp_{\bchi_{1}}B_{1}=\pp_{\bchi_{2}}B_{1}, \\
i\pp_{t}f&=&-\pp_{g}B_{1},&  \mbox{} \qquad & i\pp_{t}g&=&-\pp_{f}B_{1},
\end{array}
\end{equation}
%----------------------------------------------------------------------------------------
%	MATERIALS AND METHODS
%----------------------------------------------------------------------------------------
where the defect potentials $B_0$ and $B_1$ are given given by \cite{FLZ2},
\begin{eqnarray}
B_0 &=& m\s\left[ \cosh\phi_+-\cosh\vph_+\right]+\frac{m}{\s}\left[\cosh\phi_--\cosh\vph_-\right],\\[0.2cm]
B_1&=& i\sqrt{\frac{m\s}{2}}\left[\cosh\left(\frac{\ph_++\vph_+}{2}\right)f(\p_+-\chi_+)+ \cosh\left(\frac{\ph_+-\vph_+}{2}\right)g(\p_++\chi_+)\right]\non\\
&&\!\!- i\sqrt{\frac{m}{2\s}}\left[\cosh\left(\frac{\ph_--\vph_-}{2}\right)f(\bp_-+\bchi_-)+\cosh\left(\frac{\ph_-+\vph_-}{2}\right)g(\bp_--\bchi_-)\right],
\er
where we have denoted $\ph_\pm=\ph_1\pm\ph_2,\ \p_\pm=\p_1\pm\p_2$, $\chi_\pm=\chi_1\pm\chi_2$ for the others fields the notation is similar, and $\s$ is a free parameter associated with the defect. \\
In next section, we will perform the fusing of  two type-I defects placed at different points in order to construct a type-II defect for the $\cn=2$ sshG model.

\section{Fusing Defects}
Let us introduce two type-I defects in the $\cn=2$ sshG model, one  located at $x=0$, and a second one located  at $x=x_0$ 
\begin{figure}[h]
\begin{center}
\includegraphics[width=0.7\linewidth]{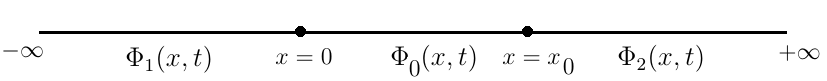}
\end{center}
\caption{\label{fig2}Fusing defects.}
\end{figure}
where $\Phi_1(\ph_1,\p_1,\bp_1,\vph_1,\chi_1,\bchi_1)$ is a set of fields in the region $x<0$, $\Phi_0(\ph_0,\p_0,\bp_0,\vph_0,\chi_0,\bchi_0)$ is the correspondingly set of fields for the region $0<x<x_0$ and $\Phi_2(\ph_2,\p_2,\bp_2,\vph_2,\chi_2,\bchi_2)$ in the $x>0$.\\
Then, the Lagrangian density describing this system can be written as,
\br\label{lagrangian}
{\mathcal L} &=& \theta(-x)  {\mathcal L}_1  +\delta(x){\mathcal L}_{D_1} + \theta(x)\theta(x_0-x){\mathcal L}_0-\delta(x-x_0) {\mathcal L}_{D_2}+
\theta(x-x_0) {\mathcal L}_2, 
\er
where the two type-I defect Lagrangian densities ${\cl}_{D_k}$ at $x=0$ $(k=1)$, and $x=x_0$ $(k=2)$, are given by
\br
{\mathcal L}_{D_k} &=&\frac{1}{2}[\phi_0\partial_t\phi_k-\phi_k\partial_t\phi_0]-\frac{1}{2}[\vph_0\partial_t\vph_k-\vph_k\partial_t\vph_0]+i(\bchi_k\bchi_0+\chi_k\chi_0)\non\\&&- i(\bp_k\bp_0+\p_k\p_0)-(-1)^{k}\left(\frac{i}{2}f_k\pp_tg_{k}+\frac{i}{2}g_k\pp_tf_k+B_1^{(k)}+B_{0}^{(k)}\right),
\er
with the defect potentials
\begin{eqnarray}
B_0^{(k)} &=& m\s_k\left[ \cosh(\phi_0+\phi_k)-\cosh(\vph_0+\vph_k)\right]+\frac{m}{\s_k}\left[\cosh(\phi_0-\phi_k)-\cosh(\vph_0-\vph_k)\right],\qquad \mbox{}\\
B_1^{(k)}&=& i\sqrt{\frac{m\s_k}{2}}\cosh\left(\frac{\ph_k+\ph_0+\vph_0+\vph_k}{2}\right)f_k(\p_0+\p_k-\chi_0-\chi_k)\non\\
&& +i\sqrt{\frac{m\s_k}{2}}\cosh\left(\frac{\ph_k+\ph_0-\vph_k-\vph_0}{2}\right)g_k(\p_k+\p_0+\chi_0+\chi_k)\non\\
&& -i(-1)^{k}\sqrt{\frac{m}{2\s_k}}\cosh\left(\frac{\ph_0-\ph_k+\vph_k-\vph_0}{2}\right)f_k(\bp_0-\bp_k+\bchi_0-\bchi_k)\non\\
&& -i(-1)^{k}\sqrt{\frac{m}{2\s_k}}\cosh\left(\frac{\ph_0-\ph_k+\vph_0-\vph_k}{2}\right)g_k(\bp_0-\bp_k-\bchi_0+\bchi_k),
\er
where $\s_k$, with $k = 1, 2$ are two free parameters associated two each defect. Thus for each defect we can write the following equations of motion\\
for $x=0:$
\begin{equation}\label{defect 0}
\begin{array}{r c l c r c l}\pp_{x}\phi_{1}-\pp_{t}\phi_{0}&=&-\pp_{\phi_{1}}(B_{0}^{(1)}+B_{1}^{(1)}),&  \mbox{} \qquad & \pp_{x}\vph_{1}-\pp_{t}\vph_{0}&=&\pp_{\vph_{1}}(B_{0}^{(1)}+B_{1}^{(1)}),\\
\pp_{x}\phi_{0}-\pp_{t}\phi_{1}&=&\pp_{\phi_{0}}(B_{0}^{(1)}+B_{1}^{(1)}),&  \mbox{} \qquad &\pp_{x}\vph_{0}-\pp_{t}\vph_{1}&=&-\pp_{\vph_{0}}(B_{0}^{(1)}+B_{1}^{(1)}), \\
i(\p_{1}-\p_{0})&=&-\pp_{\p_{1}}B_{1}^{(1)}=-\pp_{\p_{0}}B_{1}^{(1)},&  \mbox{} \qquad &i(\chi_{1}-\chi_{0})&=&\pp_{\chi_{1}}B_{1}^{(1)}=\pp_{\chi_{0}}B_{1}^{(1)},\\
i(\bp_{1}+\bp_{0})&=&\pp_{\bp_{1}}B_{1}^{(1)}=-\pp_{\bp_{0}}B_{1}^{(1)},&  \mbox{} \qquad &i(\bchi_{1}+\bchi_{0})&=&-\pp_{\bchi_{1}}B_{1}^{(1)}=\pp_{\bchi_{0}}B_{1}^{(1)}, \\
i\pp_{t}f_1&=&-\pp_{g_1}B_{1}^{(1)},&  \mbox{} \qquad & i\pp_{t}g_1&=&-\pp_{f_1}B_{1}^{(1)},
\end{array}
\end{equation}
and for $x=x_0:$
\begin{equation}\label{defect x0}
\begin{array}{r c l c r c l}\pp_{x}\phi_{0}-\pp_{t}\phi_{2}&=&-\pp_{\phi_{0}}(B_{0}^{(2)}+B_{1}^{(2)}),&  \mbox{} \qquad & \pp_{x}\vph_{0}-\pp_{t}\vph_{2}&=&\pp_{\vph_{0}}(B_{0}^{(2)}+B_{1}^{(2)}),\\
\pp_{x}\phi_{2}-\pp_{t}\phi_{0}&=&\pp_{\phi_{2}}(B_{0}^{(2)}+B_{1}^{(2)}),&  \mbox{} \qquad &\pp_{x}\vph_{2}-\pp_{t}\vph_{0}&=&-\pp_{\vph_{2}}(B_{0}^{(2)}+B_{1}^{(2)}), \\
i(\p_{0}-\p_{2})&=&-\pp_{\p_{0}}B_{1}^{(2)}=-\pp_{\p_{2}}B_{1}^{(2)},&  \mbox{} \qquad &i(\chi_{0}-\chi_{2})&=&\pp_{\chi_{0}}B_{1}^{(2)}=\pp_{\chi_{2}}B_{1}^{(2)},\\
i(\bp_{2}+\bp_{0})&=&\pp_{\bp_{0}}B_{1}^{(2)}=-\pp_{\bp_{2}}B_{1}^{(2)},&  \mbox{} \qquad &i(\bchi_{2}+\bchi_{0})&=&-\pp_{\bchi_{0}}B_{1}^{(2)}=\pp_{\bchi_{2}}B_{1}^{(2)}, \\
i\pp_{t}f_2&=&-\pp_{g_2}B_{1}^{(2)},&  \mbox{} \qquad & i\pp_{t}g_2&=&-\pp_{f_2}B_{1}^{(2)},
\end{array}
\end{equation}
Now taking the limit $x_0\to 0$ in the Lagrangian density (\ref{lagrangian}), the bulk Langrangian term $\mathcal{L}_0$ vanishes, and then the  resulting Lagrangian density for fused defect becomes of the form of eq. (\ref{fused}) with ${\mathcal L}_D={\mathcal L}_{D{1}}-{\mathcal L}_{D{2}}$, namely,
\br
{\mathcal L}_D &=& \frac{1}{2}(\phi_0\partial_t\phi_--\phi_-\partial_t\phi_0)-i(\bp_-\bp_0+\p_-\p_0)+\frac{i}{2}(f_1\pp_tg_{1}+f_{2}\pp_tg_2+g_1\pp_tf_{1}+g_2\pp_tf_{2})\non\\
&&-\frac{1}{2}(\vph_0\partial_t\vph_--\vph_-\partial_t\vph_0)+i(\bchi_-\bchi_0+\chi_-\chi_0)+B_0^{(1)}+B_0^{(2)}+B_1^{(1)}+B_1^{(2)}.
\er
%\subsection*{Bosonic Part}
We note that the fields of the bulk Langrangian term $\mathcal{L}_0$ only contribute to the total defect Lagrangian at $x=0$, and become auxiliary fields.\\
The fused bosonic potential for $\cn=2$ sshG model $B_0=B_0^{(1)}+B_0^{(2)}=B_0^++B_0^-$, is a combination of two $\cn=1$ potentials  previously obtained in \cite{Nathaly}, and it can be written explicitly as follows,
\br
B_0^{+} &=&\frac{m}{2}\left[ e^{\(\frac{\phi_+}{2}+\phi_0\)}\left(\s_1e^{\frac{\phi_-}{2}}+\s_2e^{-\frac{\phi_-}{2}}\) + e^{-\(\frac{\phi_+}{2}+\phi_0\)}\left(\s_1e^{-\frac{\phi_-}{2}}+\s_2e^{\frac{\phi_-}{2}}\)\right.\non\\
&&\quad \,\,-\left. e^{\(\frac{\vph_+}{2}+\vph_0\)}\left(\s_1e^{\frac{\vph_-}{2}}+\s_2e^{-\frac{\vph_-}{2}}\) - e^{-\(\frac{\vph_+}{2}+\vph_0\)}\left(\s_1e^{-\frac{\vph_-}{2}}+\s_2e^{\frac{\vph_-}{2}}\)\right],\label{4.20}\\[0.1cm]
B_0^{-} &=&\frac{m}{2}\left[e^{\(\frac{\phi_+}{2}-\phi_0\)}\left(\frac{1}{\s_1} e^{\frac{\phi_-}{2}}+\frac{1}{\s_2}e^{-\frac{\phi_-}{2}}\) + e^{-\(\frac{\phi_+}{2}-\phi_0\)}\left(\frac{1}{\s_1}e^{-\frac{\phi_-}{2}}+\frac{1}{\s_2}e^{\frac{\phi_-}{2}}\)\right.\non\\
&&\quad \,\,-\left. e^{\(\frac{\vph_+}{2}-\vph_0\)}\left(\frac{1}{\s_1} e^{\frac{\vph_-}{2}}+\frac{1}{\s_2}e^{-\frac{\vph_-}{2}}\) - e^{-\(\frac{\vph_+}{2}-\vph_0\)}\left(\frac{1}{\s_1}e^{-\frac{\vph_-}{2}}+\frac{1}{\s_2}e^{\frac{\vph_-}{2}}\) \right].\label{4.20}
\er
%\subsection*{Fermionic Part}
For the fermionic part we need to use the equations of motion (\ref{mov geral def}) for each region, in order to eliminate the auxiliary fields $\bp_0,\p_0,\chi_0,\bchi_0$, we get
\br
\p_0&=& \frac{\p_+}{2}-\sqrt{\frac{m\s_1}{2}}\left[\pp_{\ph_1}u_1^{+}f_1+\pp_{\ph_1}u_1^{-}g_1\right]+\sqrt{\frac{m\s_2}{2}}\left[\pp_{\ph_2}u_2^{+}f_2+\pp_{\ph_2}u_2^{-}g_2\right]\\
\chi_0&=& \frac{\chi_+}{2}-\sqrt{\frac{m\s_1}{2}}\left[\pp_{\ph_1}u_1^{+}f_1-\pp_{\ph_1}u_1^{-}g_1\right]+\sqrt{\frac{m\s_2}{2}}\left[\pp_{\ph_2}u_2^{+}f_2-\pp_{\ph_2}u_2^{-}g_2\right]\\
\bp_0&=& -\frac{\bp_+}{2}+\sqrt{\frac{m}{2\s_1}}\left[\pp_{\ph_1}v_1^{-}f_1+\pp_{\ph_1}v_1^{+}g_1\right]+\sqrt{\frac{m}{2\s_2}}\left[\pp_{\ph_2}v_2^{-}f_2+\pp_{\ph_2}v_2^{+}g_2\right]\\
\bchi_0&=& -\frac{\bchi_+}{2}-\sqrt{\frac{m}{2\s_1}}\left[\pp_{\ph_1}v_1^{-}f_1-\pp_{\ph_1}v_1^{+}g_1\right]-\sqrt{\frac{m}{2\s_2}}\left[\pp_{\ph_2}v_2^{-}f_2-\pp_{\ph_2}v_2^{+}g_2\right]
\er
where we define the functions
\br
u_k^{\pm} =  \sinh\left(\frac{(\ph_k+\ph_0)\pm(\vph_k+\vph_0)}{2}\right),\quad
v_k^\pm =\sinh\left(\frac{(\ph_k-\ph_0)\pm(\vph_k-\vph_0)}{2}\right)
\er
Then noting that
\br
i(\chi_-\chi_0-\p_-\p_0)&=& \frac{i}{2}(\chi_-\chi_+-\p_-\p_+)-im\sqrt{\s_1\s_2}\left[\pp_{\ph_1}u_1^{+}\pp_{\ph_2}u_2^{-}f_1g_2+\pp_{\ph_1}u_1^{-}\pp_{\ph_2}u_2^{+}g_1f_2\right],\non\\
i(\bchi_-\bchi_0-\bp_-\bp_0)&=& -\frac{i}{2}(\bchi_-\bchi_+-\bp_-\bp_+)-\frac{im}{\sqrt{\s_1\s_2}}\left[\pp_{\ph_1}v_1^{-}\pp_{\ph_2}v_2^{+}f_1g_2+\pp_{\ph_1}v_1^{+}\pp_{\ph_2}v_2^{-}g_1f_2\right],\non\\
\er
we find that the fermionic part of the fused defect Lagrangian is given by,
\br
\mathcal{L}_D\Big|_{fermion}&=& i(\bp_1\bp_2-\p_1\p_2)-i(\bchi_1\bchi_2-\chi_1\chi_2)+\frac{i}{2}(f_1\pp_tg_{1}+f_{2}\pp_tg_2+g_1\pp_tf_{1}+g_2\pp_tf_{2})\non\\&&+B_1^{+}+B_1^{-},
\er
where
\br
B_{1}^{+}&=&\frac{i}{2}\sqrt{\frac{m}{2}}\left[e^{-\big(\frac{\phi_++\vph_+}{4}+\frac{\phi_0+\vph_0}{2}\big)}\Big(\sqrt{\s_2}\, e^{\frac{\phi_-+\vph_-}{4}}f_2+\sqrt{\s_1}\, e^{-\big(\frac{\phi_-+\vph_-}{4}\big)}f_1\Big)\right.\non\\
&&\qquad \, +\left. e^{\big(\frac{\phi_++\vph_+}{4}+\frac{\phi_0+\vph_0}{2}\big)}\Big(\sqrt{\s_2}\, e^{-\big(\frac{\phi_-+\vph_-}{4}\big)}f_2+\sqrt{\s_1}\, e^{\frac{\phi_-+\vph_-}{4}}f_1\Big)\right]\!(\p_+-\chi_+)\non\\
&&+\frac{i}{2}\sqrt{\frac{m}{2}}\left[e^{-\big(\frac{\phi_+-\vph_+}{4}+\frac{\phi_0-\vph_0}{2}\big)}\Big(\sqrt{\s_1}\, e^{-\big(\frac{\phi_--\vph_-}{4}\big)}g_1+\sqrt{\s_2}\, e^{\frac{\phi_--\vph_-}{4}}g_2\Big)\right.\non\\
&&\qquad \,\,\, \,\,+\left. e^{\big(\frac{\phi_+-\vph_+}{4}+\frac{\phi_0-\vph_0}{2}\big)}\Big(\sqrt{\s_1}\, e^{\frac{\phi_--\vph_-}{4}}g_1+\sqrt{\s_2}\, e^{-\big(\frac{\phi_--\vph_-}{4}\big)}g_2\Big)\right]\!(\p_++\chi_+)\non\\
&&+ \frac{im}{2}\sqrt{\s_1\s_2}\Big(\cosh\left(\ph_0+\frac{\ph_+}{2}+\frac{\vph_-}{2}\right)+\cosh\left(\vph_0+\frac{\vph_+}{2}+\frac{\ph_-}{2}\right)\Big)f_1g_2\non\\
&&+ \frac{im}{2}\sqrt{\s_1\s_2}\Big(\cosh\left(\ph_0+\frac{\ph_+}{2}-\frac{\vph_-}{2}\right)+\cosh\left(\vph_0+\frac{\vph_+}{2}-\frac{\ph_-}{2}\right)\Big)g_1f_2,\label{4.17}
\er
and
\br
B_{1}^{-}&=&-\frac{i}{2}\sqrt{\frac{m}{2}}\left[e^{-\big(\frac{\phi_++\vph_+}{4}-\frac{\phi_0+\vph_0}{2}\big)}\Big(\frac{1}{\sqrt{\s_1}}e^{-\big(\frac{\phi_-+\vph_-}{4}\big)}g_1-\frac{1}{\sqrt{\s_2}}e^{\frac{\phi_-+\vph_-}{4}}g_2\Big)\right.\non\\
&&\qquad \,\,\,\,\,\,+\left. e^{\big(\frac{\phi_++\vph_+}{4}-\frac{\phi_0+\vph_0}{2}\big)}\Big(\frac{1}{\sqrt{\s_1}}e^{\frac{\phi_-+\vph_-}{4}}g_1-\frac{1}{\sqrt{\s_2}}e^{-\big(\frac{\phi_-+\vph_-}{4}\big)}g_2\Big)\right]\!(\bp_+-\bchi_+)\non\\
&&-\frac{i}{2}\sqrt{\frac{m}{2}}\left[e^{-\big(\frac{\phi_+-\vph_+}{4}-\frac{\phi_0-\vph_0}{2}\big)}\Big(\frac{1}{\sqrt{\s_1}}e^{-\big(\frac{\phi_--\vph_-}{4}\big)}f_1-\frac{1}{\sqrt{\s_2}}e^{\frac{\phi_--\vph_-}{4}}f_2\Big)\right.\non\\
&&\qquad \,\,\,\,\,\,+\left.
e^{\big(\frac{\phi_+-\vph_+}{4}-\frac{\phi_0-\vph_0}{2}\big)}\Big(\frac{1}{\sqrt{\s_1}}e^{\frac{\phi_--\vph_-}{4}}f_1-\frac{1}{\sqrt{\s_2}}e^{-\big(\frac{\phi_--\vph_-}{4}\big)}f_2\Big)\right]\!(\bp_++\bchi_+)\non\\
&&+ 
\frac{im}{2\sqrt{\s_1\s_2}}\Big(\cosh\left(\ph_0-\frac{\ph_+}{2}+\frac{\vph_-}{2}\right)+\cosh\left(\vph_0-\frac{\vph_+}{2}+\frac{\ph_-}{2}\right)\Big)f_1g_2\non\\
&&+ \frac{im}{2\sqrt{\s_1\s_2}}\Big(\cosh\left(\ph_0-\frac{\ph_+}{2}-\frac{\vph_-}{2}\right)+\cosh\left(\vph_0-\frac{\vph_+}{2}-\frac{\ph_-}{2}\right)\Big)g_1f_2. \label{4.18}
\er
Finally the type-II defect Lagrangian density can be expressed as follows,
\br
{\mathcal L}_D &=& \frac{1}{2}(\phi_0\partial_t\phi_--\phi_-\partial_t\phi_0)-\frac{1}{2}(\vph_0\partial_t\vph_--\vph_-\partial_t\vph_0)+i(\bp_1\bp_2-\p_1\p_2)-i(\bchi_1\bchi_2-\chi_1\chi_2)\non\\
&&+\frac{i}{2}(f_1\pp_tg_{1}+f_{2}\pp_tg_2+g_1\pp_tf_{1}+g_2\pp_tf_{2}) +B_0^{+}+B_0^{-} +B_1^{+} +B_1^{-}.
\er
Then, the corresponding type-II defects conditions for the $\cn=2$ sshG model at $x=0$ are,
\begin{equation}\label{defect}
\begin{array}{rcl c rcl}
\pp_x\ph_1-\pp_t\ph_0 &=& -\pp_{\ph_1}(B_0+B_1),& \mbox{}\quad &\pp_x\ph_2-\pp_t\ph_0 &=& \pp_{\ph_2}(B_0+B_1),\\[0.1cm]
\pp_x\vph_1-\pp_t\vph_0 &=& \pp_{\vph_1}(B_0+B_1),& \mbox{}\quad &\pp_x\vph_2-\pp_t\vph_0 &=& -\pp_{\vph_2}(B_0+B_1),\\[0.1cm]
\pp_t(\ph_1-\ph_2)&=& -\pp_{\ph_0}(B_0+B_1),& \mbox{}\quad &\pp_t(\vph_1-\vph_2)&=& \pp_{\vph_0}(B_0+B_1),\\[0.1cm]
i(\p_1-\p_2)&=& -\pp_{\p_1}B_1=-\pp_{\p_2}B_1,& \mbox{}\quad &i(\bp_1-\bp_2)&=& \pp_{\bp_1}B_1=\pp_{\bp_2}B_1,\\[0.1cm]
i(\chi_1-\chi_2)&=& \pp_{\chi_1}B_1=\pp_{\chi_2}B_1,& \mbox{}\quad &i(\bchi_1-\bchi_2)&=& -\pp_{\bchi_1}B_1=-\pp_{\bchi_2}B_1,\\[0.1cm]
i\pp_tg_1&=&-\pp_{f_1}B_1,& \mbox{}\quad &i\pp_tf_1&=&-\pp_{g_1}B_1,\\[0.1cm]
i\pp_tg_2&=&-\pp_{f_2}B_1,& \mbox{}\quad &i\pp_tf_2&=&-\pp_{g_2}B_1.
\end{array}
\end{equation}
The explicit form of the B\"acklund transformation for $\cn=2$ sshG model is presented in \mbox{appendix A.}

%%%%%%%%%%%%%%%%%%%%%%%%%%%%%%%%%%%%%%%%%%%%%%%%%%%

\section{Conservation of the momentum and energy}
In this section, we will discuss the modified conserved momentum and energy. Let us consider first the total canonical momentum, which is given by the following contributions
\br
P=\int_{-\infty}^{0}dx\, {\cal P}_1+\int_{0}^{+\infty}dx\, {\cal P}_2,
\er
with
\br
{\cal P}_p=\pp_t\ph_p\pp_x\ph_p-\pp_t\vph_p\pp_x\vph_p-i(\p_p\pp_x\p_p+\bp_p\pp_x\bp_p)+i(\chi_p\pp_x\chi_p+\bchi_p\pp_x\bchi_p),\quad p=1,2.
\er
% % % % %
Using the bulk equations (\ref{mov geral 1}), we can write the time derivative of momentum as
\br
\frac{dP}{dt}&=&\Big[ \frac{1}{2}(\pp_x\ph_1)^2+\frac{1}{2}(\pp_t\ph_1)^2-\frac{1}{2}(\pp_x\vph_1)^2-\frac{1}{2}(\pp_t\vph_1)^2-i(\p_1\pp_t\p_1+\bp_1\pp_t\bp_1)\non\\&&\, + i(\chi_1\pp_t\chi_1+\bchi_1\pp_t\bchi_1)-\frac{1}{2}(\pp_x\ph_2)^2-\frac{1}{2}(\pp_t\ph_2)^2+\frac{1}{2}(\pp_x\vph_2)^2+\frac{1}{2}(\pp_t\vph_2)^2\non\\&&\,+
i(\p_2\pp_t\p_2+\bp_2\pp_t\bp_2)-i(\chi_2\pp_t\chi_2+\bchi_2\pp_t\bchi_2)-V_1+V_2-W_1+W_2\Big]_{x=0}.\label{4.25}
\er
Now, from the explicit form of the defect potentials $B_0=B_0^{+}+B_0^-,\ B_1=B_1^{+}+B_1^-$ given in eqs. (\ref{4.17})--(\ref{4.20}), and the defect conditions (\ref{defect}),  we find the following set of relations,
\br 
\pp_{\p_-}B_1\!&=&\!\pp_{\bp_-}B_1\,=\,\pp_{\chi_-}B_1\,=\,\pp_{\bchi_-}B_1\,=\,0,\non\\
\pp_{\p_+}B_1^-\!&=&\!\pp_{\bp_+}B_1^+=\pp_{\chi_+}B_1^-=\pp_{\bchi_+}B_1^+=0,
\er
and 
\begin{equation}
\begin{array}{rcl c rcl}
\pp_{\ph_0}B_0^+&=& 2\pp_{\ph_+}B_0^+,&\mbox{}\,\,\,& \pp_{\ph_0}B_1^+ &=& 2\pp_{\ph_+}B_1^+ ,\\ \pp_{\ph_0}B_0^-&=&-2\pp_{\ph_+}B_0^-,&\mbox{}\,\,\,&\pp_{\ph_0}B_1^-&=&-2\pp_{\ph_+}B_1^-,\\
\pp_{\vph_0}B_0^+&=& 2\pp_{\vph_+}B_0^+,&\mbox{}\,\,\,& \pp_{\vph_0}B_1^+&=& 2\pp_{\vph_+}B_1^+,\\ \pp_{\vph_0}B_0^-&=&-2\pp_{\vph_+}B_0^-,&\mbox{}\,\,\,&\pp_{\vph_0}B_1^-&=&-2\pp_{\vph_+}B_1^-.
\end{array}
\end{equation}
Then, by using the above relations and the defect conditions (\ref{defect}), the equation (\ref{4.25})  takes the following form,
\br
\frac{dP}{dt}&=&\Big[2\pp_{\ph_+}B_0\pp_{\ph_-}B_0-2\pp_{\vph_+}B_0\pp_{\vph_-}B_0+2\pp_{\ph_+}B_1\pp_{\ph_-}B_1-2\pp_{\vph_+}B_1\pp_{\vph_-}B_1+2\pp_{\ph_-}B_0\pp_{\ph_+}B_1
\non\\&&\,\,+2\pp_{\ph_+}B_0\pp_{\ph_-}B_1-2\pp_{\vph_-}B_0\pp_{\vph_+}B_1-2\pp_{\vph_+}B_0\pp_{\vph_-}B_1-2\pp_t\ph_0\pp_{\ph_+}(B_0+B_1)\non\\
&&\,\,-2\pp_t\vph_0\pp_{\vph_+}(B_0+B_1)-\frac{1}{2}\pp_t\ph_+\pp_{\ph_0}(B_0+B_1)-\frac{1}{2}\pp_t\vph_+\pp_{\vph_0}(B_0+B_1)\non\\
&&\,\,-\pp_t\p_+\pp_{\p_+}B_1+
\pp_t\bp_+\pp_{\bp_+}B_1-\pp_t\chi_+\pp_{\chi_+}B_1+\pp_t\bchi_+\pp_{\bchi_+}B_1-V_1+V_2-W_1+W_2\non\\&&\,\,+i\pp_t(\p_1\p_2+\bp_1\bp_2-\chi_1\chi_2-\bchi_1\bchi_2)\Big]_{x=0},
\er
where the right-hand-side of the above equation becomes a total time derivative since the defect potentials $B_0^{\pm}$ and $B_1^{\pm}$ satisfy the following Poisson-bracket-like relations,
\br
V_1-V_2&=& 2(\pp_{\ph_0}B_0^+\pp_{\ph_-}B_0^--\pp_{\ph_0}B_0^-\pp_{\ph_-}B_0^+-
\pp_{\vph_0}B_0^+\pp_{\vph_-}B_0^-+\pp_{\vph_0}B_0^-\pp_{\vph_-}B_0^+),\label{4.28}\\[0.2cm]
W_1-W_2&=& \,\,2(\pp_{\ph_0}B_1^+\pp_{\ph_-}B_0^--\pp_{\ph_0}B_1^-\pp_{\ph_-}B_0^++\pp_{\ph_0}B_0^+\pp_{\ph_-}B_1^--\pp_{\ph_0}B_0^-\pp_{\ph_-}B_1^+)\non\\
&&\!\!\!-2(\pp_{\vph_0}B_1^+\pp_{\vph_-}B_0^--\pp_{\vph_0}B_1^-\pp_{\vph_-}B_0^++\pp_{\vph_0}B_0^+\pp_{\vph_-}B_1^--\pp_{\vph_0}B_0^-\pp_{\vph_-}B_1^+)\non\\
&&\!\!\!+2i(\pp_{f_1}B_1^-\pp_{g_1}B_1^+-\pp_{f_1}B_1^+\pp_{g_1}B_1^-+\pp_{f_2}B_1^-\pp_{g_2}B_1^+-\pp_{f_2}B_1^+\pp_{g_2}B_1^-),
\er
together with the constraint,
\br
\pp_{\ph_0}B_1^+\pp_{\ph_-}B_1^--\pp_{\ph_0}B_1^-\pp_{\ph_-}B_1^+-
\pp_{\vph_0}B_1^+\pp_{\vph_-}B_1^-+\pp_{\vph_0}B_1^-\pp_{\vph_-}B_1^+=0.
\er%
Then, we get that the modified conserved momentum can be written in a simple form, 
\br
\mathcal{P}&=& P+\Big[B_1^++B_0^+-B_1^--B_0^-+i(-\p_1\p_2-\bp_1\bp_2+\chi_1\chi_2+\bchi_1\bchi_2)\Big]_{x=0}.
\er
Now, let us consider the total energy
\br
E=\int_{-\infty}^{0}dx\, {\cal E}_1+\int_{0}^{+\infty}dx\, {\cal E}_2,
\er
where
\br
{\cal E}_p&=&\frac{1}{2}(\pp_x\ph_p)^2+\frac{1}{2}(\pp_t\ph_p)^2-\frac{1}{2}(\pp_x\vph_p)^2-\frac{1}{2}(\pp_t\vph_p)^2-i(\p_p\pp_x\p_p-\bp_p\pp_x\bp_p)\non\\&&+ i(\chi_p\pp_x\chi_p-\bchi_p\pp_x\bchi_p)+V_p+W_p.
\er
We can find its time-derivative in the same way as before by using the bulk equations (\ref{mov geral 1}). The result reads
\br
\frac{dE}{dt}&=&\Big[\pp_t\ph_1\pp_x\ph_1-\pp_t\vph_1\pp_x\vph_1-i(\p_1\pp_t\p_1-\bp_1\pp_t\bp_1)+i(\chi_1\pp_t\chi_1-\bchi_1\pp_t\bchi_1)\non\\&&-
\pp_t\ph_2\pp_x\ph_2+\pp_t\vph_2\pp_x\vph_2+i(\p_2\pp_t\p_2-\bp_2\pp_t\bp_2)-i(\chi_2\pp_t\chi_2-\bchi_2\pp_t\bchi_2)\Big]_{x=0}.
\er
Then using the defect conditions (\ref{defect}) and the defect potentials (\ref{4.17})--(\ref{4.20}), we find that the modified conserved energy is given by
\br
\mathcal{E} = E +\Big[B_0+B_1+i(\bp_1\bp_2-\p_1\p_2+\chi_1\chi_2-\bchi_1\bchi_2)\Big]_{x=0}.
\er

%%%%%%%%%%%%%%%%%%%%%%%%%%%%%%%%%%%%%%%%%%%%%%%%%%%%%%%%%%%%%%%

%%%%%%%%%%%%%%%%%%%%%%%%%%%%%%%%%%%%%%%%%%%%%%%%%%%%%%%%%%%%%%%

\section{Modified conserved supercharges}

We have seen that the bulk theory  action is invariant under susy
transformation (\ref{susy}), and it was explicitly shown for $\d_1$ projection. However, this is not necessarily true for the defect theory, and therefore we should show that the presence of the defect will not destroy the supersymmetry of the bulk theory. Let us compute the defect contribution for $Q_1$. By introducing the defect at $x = 0$, we have
\br
Q_1 &=&\int_{-\infty}^{0} dx \Big[2i\p_1\pp_+\ph_1 + 2i\chi_1\pp_+\vph_1 -2im \bp_1\sinh\ph_1 \cosh\vph_1 +2im\bchi_1\sinh\vph_1\cosh\ph_1 \Big]\non\\
&& + \int_{0}^{\infty} dx \Big[2i\p_2\pp_+\ph_2+ i\chi_2\pp_+\vph_2 -2im \bp_2\sinh\ph_2 \cosh\vph_2+2im\bchi_2\sinh\vph_2\cosh\ph_2 \Big].\,\,\, \mbox{}
\er
Now, by taking the time-derivative respectively, we get
\br
\frac{dQ_1}{dt}&=&\Big[2i\p_1\pp_+\ph_1 + 2i\chi_1\pp_+\vph_1 +2im \bp_1\sinh\ph_1 \cosh\vph_1 -2im\bchi_1\sinh\vph_1\cosh\ph_1\Big]_{x=0}\non\\
&&\!\!\!\!-\Big[2i\p_2\pp_+\ph_2 + 2i\chi_2\pp_+\vph_2 +2im \bp_2\sinh\ph_2 \cosh\vph_2 -2im\bchi_2\sinh\vph_2\cosh\ph_2\Big]_{x=0}.\quad \mbox{}
\er
Using the defect conditions (\ref{defect}), we get
\br
\frac{dQ_1}{dt} &=&\!\Bigg[ \,\,\,i\psi_- \pp_t\left( \frac{\ph_+}{2}+\ph_0 \right) -i\psi_-\,\pp_{\ph_-}(B_0+B_1) -i\psi_+\,\pp_{\ph_0}(B_0^++B_1^+) \non \\
&&+ i\chi_- \,\pp_t\left( \frac{\vph_+}{2}+\vph_0 \right)+i\chi_-\pp_{\vph_-}(B_0+B_1) +i\chi_+\,\pp_{\vph_0}(B_0^++B_1^+)\non \\
&&-im (\bp_++\bp_-)\sinh\Big(\frac{\ph_++\ph_-}{2}\Big)\cosh\Big(\frac{\vph_++\vph_-}{2}\Big)\non\\
&& +im(\bp_+-\bp_-)\sinh\Big(\frac{\ph_+-\ph_-}{2}\Big)\cosh\Big(\frac{\vph_+-\vph_-}{2}\Big)\non \\
&&+im (\bchi_++\bchi_-)\sinh\Big(\frac{\vph_++\vph_-}{2}\Big)\cosh\Big(\frac{\ph_++\ph_-}{2}\Big)\non\\
&& -im(\bchi_+-\bchi_-)\sinh\Big(\frac{\vph_+-\vph_-}{2}\Big)\cosh\Big(\frac{\ph_+-\ph_-}{2}\Big)\Bigg]_{x=0}.
\er
Now, by making use of the defect conditions intensively, we find after some algebra that
the right-hand-side of the equation becomes a total time-derivative, and then the modified conserved supercharge can be written in ${\cal Q}_1=Q_1+Q_{{\scriptsize D}_1}$, with the defect contribution given by the following expression,
\br 
Q_{{\tiny D}_1}= \sum\limits_{k=1}^{2} -i\sqrt{2m\s_k}\Big( u_k^+ f_k+ u_k^-g_k \Big)_{x=0}.
\er
where we have introduced the function,
\br
u_k^{\pm} =  \sinh\left(\frac{(\ph_k+\ph_0)\pm(\vph_k+\vph_0)}{2}\right).
\er
Analogously, we can find the remaining modified conserved supercharges, 
\br
{\cal Q}_2 =Q_2+ Q_{D_2}, \qquad  \widebar{\cal Q}_1 =\widebar{Q}_1+ \widebar{Q}_{D_1}, \qquad \widebar{\cal Q}_2 =\widebar{Q}_2+ \widebar{Q}_{D_2}, 
\er
with the corresponding defect contributions given by,
\br 
Q_{{\tiny D}_2}&=& \sum\limits_{k=1}^{2} -i\sqrt{2m\s_k}\Big( u_k^+ f_k- u_k^- g_k \Big)_{x=0},\\
{\widebar Q}_{{\tiny D}_1}&=& \sum\limits_{k=1}^{2} \frac{i\sqrt{2m}(-1)^k}{\sqrt{\s_k}}\Big( v_k^- f_k+ v_k^+ g_k \Big)_{x=0},\\
{\widebar Q}_{{\tiny D}_2}&=& \sum\limits_{k=1}^{2} \frac{i\sqrt{2m}(-1)^{k-1}}{\sqrt{\s_k}}\Big(v_k^- f_k-v_k^+ g_k \Big)_{x=0},
\er
where the functions $v_k^\pm$ are defined to be,
\br
v_k^\pm =\sinh\left(\frac{(\ph_k-\ph_0)\pm(\vph_k-\vph_0)}{2}\right).
\er
The derivation of the exact form of the all modified conserved, together with the modified conserved energy and momentum,  provides a strong evidence indicating the classical integrability of the fused defect for the $\cn=2$ sshG model. A more rigorous analysis should require the derivation of the generating function of an infinite set of modified conserved quantities. That can be done following the on-shell Lax approach to derive the corresponding type-II defect ${\cal K}$-matrix for the model. From its explicit form is possible to derive an infinite set of modified conserved quantities. Alternative approaches can also be used in order to prove the involutivity of the charges, for instance the off-shell r-matrix and the multisymplectic approach as well. Some of these issues will be considered in future investigations.

\section{PT symmetry}

First of all, it can be shown that the bulk Lagrangian density and fields equations are invariant under the simultaneous transformations of parity transformation (${\bf P}$), and time reversal (${\bf T}$), namely $(x,t)\to(-x,-t)$, if the fields transform as follows,
\br
\ph(x,t)&\longrightarrow&\ph(-x,-t), \qquad \qquad \vph(x,t)\longrightarrow\vph(-x,-t), \non\\
\psi(x,t)&\longleftrightarrow& -\chi(-x,-t), \quad \qquad \,\, \bp(x,t) \longleftrightarrow-\bchi(-x,-t), \non 
\er
In addition, we notice that applying this ${\bf PT}$ transformation, the $\ep_1$-projection   of the susy transformation maps to the $\ep_2$-projection. Then, as it can be verified, by applying the ${\bf PT}$ transformation over $Q_1$ we get the second supercharge, namely $Q_2={\bf PT}Q_1$. Analogously, it happens with  $ \widebar{Q}_2={\bf PT}\widebar{Q}_1$.

Now, in the presence of a type-I defect the ${\bf PT}$ transformation relates the fields on the left-hand side to the ones on the right-hand side, and conversely. Then, to preserve the invariance under PT symmetry the fields in the respective bulk Lagrangian densities ${\cal L}_p$ should transform in the following way,
\br
\ph_1(x,t) &\longleftrightarrow& \ph_2(-x,-t), \qquad \qquad \,\,\vph_1(x,t) \longleftrightarrow \vph_2(-x,-t), \non\\
\p_1(x,t)&\longleftrightarrow &-\chi_2(-x,-t), \qquad \qquad \!\!\!\bp_1(x,t)\longleftrightarrow -\bchi_2(-x,-t).\non
\er
Consequently, the auxiliary fermionic fields $f, g$ in the defect Lagrangian should transform as,
\br
f(t)\longrightarrow f(-t), \qquad g(t)\longrightarrow -g(-t).\non 
\er
Under these field transformations it can be shown that the type-I defect equations are invariant. On the other hand,
the type-II defect Lagrangian density for the $\cn=2$ sshG model is invariant under ${\bf PT}$ transformation, if the corresponding auxiliary fields transform in the following way,
\br
\ph_0(t) &\longleftrightarrow& \ph_0(-t), \qquad \quad  \,\,\,\,\,\, \vph_0(t) \longleftrightarrow \vph_0(-t), \non \\ 
f_1(t) &\longleftrightarrow &  f_2(-t), \qquad \qquad  g_1(t) \longleftrightarrow -g_2(-t).
\er
In this case, we can verified that if this ${\bf PT}$ transformation is applied over $Q_{D_1}$ we obtain the defect contribution to the second supercharge, namely $Q_{D_2}={\bf PT}Q_{D_1}$. The same is valid for $\widebar{Q}_{D_2}={\bf PT}\widebar{Q}_{D_1}$.

The invariance of the $\cn=2$ sshG model under ${\bf PT}$ symmetry is strongly related with the description of the equation of motion in the superspace formalism. In such language, the fields appear as components of two $\cn=2$ superfields, one of them being a chiral superfield, while  the other one is anti-chiral (For more details see \cite{FLZ2}). The fact that the ${\bf PT}$ symmetry is preserved in the presence of the type-II defect, somehow suggests the possibility of describing the defect conditions in terms of superfields. In other words, there should exists a type-II B\"acklund transformation for the $\cn=2$ sshG equation consistent with the defect conditions of the fused defect. As it was shown for the $\cn=1$ sshG equations, the type-II defect conditions are equivalent to ``frozen" type-II B\"acklund transformation of the model (see appendix A).

%%%%%%%%%%%%%%%%%%%%%%%%%%%%%%%%%%%%%%%%%%%%%%%%%%%%%%%%%%%%%%%
%\newpage
\section{Final Remarks}

In this paper, we have derived a type-II integrable defect for the $\cn =2$ sshG model by using the fusing procedure. At the Lagrangian level, we have shown that
the type-II defect for this supersymmetric model can be also obtained by fusing two type-I defects located initially at different points in the $x$-axis. We have shown the conservation of the modified quantities of the energy, momentum and supercharges. Moreover, the invariance under ${\bf PT}$ symmetry was verified.

From the results obtained in this paper and those previously found in \cite{Nathaly}, it would be interesting to explore the possibility of finding  new integrable boundary conditions for the $\cn=1$ and $\cn=2$ sshG models, by performing a consistent half-line limit, specially following the reasoning of \cite{Zambon1}--\cite{Zambon2}.  

There are many others algebraic aspects related to  type-II defects  that have not been addressed in this work, like the Lax representation, the involutivity of the charges via the r-matrix approach, and the construction of the soliton solutions. These issues are expected to be addressed in future investigations.

\ack 
The authors are very grateful to the organisers of the XXIIIth International Conference on Integrable Systems and Quantum Symmetries (ISQS-23) for the opportunity to present this work. NIS,  JFG and AHZ would like to thank CNPq-Brasil for financial support.  ARA is supported by FAPESP grant 2012/13866-3.

\appendix

%%%%%%%%%%%%%%%%%%%%%%%%%%%%%%%%%%%%%%%%%%%%%%%%%%%%%%%%%%%%%%%%%%%%%%%%%%%%
\section{Type-II B\"acklund transformations for $\cn=2$ sshG model}
\label{appA}
\br
\p_-&=&\sqrt{2m}\left[\sqrt{\s_1}(\pp_{\ph_0}u_1^{+}f_1+\pp_{\ph_0}u_1^{-}g_1)+\sqrt{\s_2}(\pp_{\ph_0}u_2^{+}f_2+\pp_{\ph_0}u_2^{-}g_2)\right]\\
\bp_-&=&-\sqrt{2m}\left[\frac{1}{\sqrt{\s_1}}(\pp_{\ph_0}v_1^{+}g_1+\pp_{\ph_0}v_1^{-}f_1)-\frac{1}{\sqrt{\s_2}}(\pp_{\ph_0}v_2^{+}g_2+\pp_{\ph_0}v_2^{-}f_2)\right]\\
\chi_-&=&\sqrt{2m}\left[\sqrt{\s_1}(\pp_{\ph_0}u_1^{+}f_1-\pp_{\ph_0}u_1^{-}g_1)+\sqrt{\s_2}(\pp_{\ph_0}u_2^{+}f_2-\pp_{\ph_0}u_2^{-}g_2)\right]\\
\bchi_-&=&-\sqrt{2m}\left[\frac{1}{\sqrt{\s_1}}(\pp_{\ph_0}v_1^{+}g_1-\pp_{\ph_0}v_1^{-}f_1)-\frac{1}{\sqrt{\s_2}}(\pp_{\ph_0}v_2^{+}g_2-\pp_{\ph_0}v_2^{-}f_2)\right]\\
\pp_-g_1&=&\frac{m}{2\sqrt{\s_1\s_2}}\Big(\cosh\left(\ph_0-\frac{\ph_+}{2}+\frac{\vph_-}{2}\right)+\cosh\left(\vph_0-\frac{\vph_+}{2}+\frac{\ph_-}{2}\right)\Big)g_2\non\\&+&\sqrt{\frac{2m}{\s_1}}\pp_{\ph_0}v_1^{-}(\bp_++\bchi_+)\\
\pp_+g_1&=&-\frac{m}{2}\sqrt{\s_1\s_2}\Big(\cosh\left(\ph_0+\frac{\ph_+}{2}+\frac{\vph_-}{2}\right)+\cosh\left(\vph_0+\frac{\vph_+}{2}+\frac{\ph_-}{2}\right)\Big)g_2\non\\&-&\sqrt{2m\s_1}\pp_{\ph_0}u_1^{+}(\p_+-\chi_+)\\
\pp_-f_1&=&\frac{m}{2\sqrt{\s_1\s_2}}\Big(\cosh\left(\ph_0-\frac{\ph_+}{2}-\frac{\vph_-}{2}\right)+\cosh\left(\vph_0-\frac{\vph_+}{2}-\frac{\ph_-}{2}\right)\Big)f_2\non\\
&+&\sqrt{\frac{2m}{\s_1}}\pp_{\ph_0}v_1^{+}(\bp_+-\bchi_+)\\
\pp_+f_1&=&-\frac{m}{2}\sqrt{\s_1\s_2}\Big(\cosh\left(\ph_0+\frac{\ph_+}{2}-\frac{\vph_-}{2}\right)+\cosh\left(\vph_0+\frac{\vph_+}{2}-\frac{\ph_-}{2}\right)\Big)f_2\non\\
&-&\sqrt{2m\s_1}\pp_{\ph_0}u_1^{-}(\p_++\chi_+)\\
\pp_-g_2&=&-\frac{m}{2\sqrt{\s_1\s_2}}\Big(\cosh\left(\ph_0-\frac{\ph_+}{2}-\frac{\vph_-}{2}\right)+\cosh\left(\vph_0-\frac{\vph_+}{2}-\frac{\ph_-}{2}\right)\Big)g_1\non\\
&-&\sqrt{\frac{2m}{\s_2}}\pp_{\ph_0}v_2^{-}(\bp_++\bchi_+)\\
\pp_+g_2&=& \frac{m}{2}\sqrt{\s_1\s_2}\Big(\cosh\left(\ph_0+\frac{\ph_+}{2}-\frac{\vph_-}{2}\right)+\cosh\left(\vph_0+\frac{\vph_+}{2}-\frac{\ph_-}{2}\right)\Big)g_1\non\\
&-&\sqrt{2m\s_2}\pp_{\ph_0}u_2^{+}(\p_+-\chi_+)\\
\pp_-f_2&=&-\frac{m}{2\sqrt{\s_1\s_2}}\Big(\cosh\left(\ph_0-\frac{\ph_+}{2}+\frac{\vph_-}{2}\right)+\cosh\left(\vph_0-\frac{\vph_+}{2}+\frac{\ph_-}{2}\right)\Big)f_1\non\\
&-&\sqrt{\frac{2m}{\s_2}}\pp_{\ph_0}v_2^{+}(\bp_+-\bchi_+)\\
\pp_+f_2&=& \frac{m}{2}\sqrt{\s_1\s_2}\Big(\cosh\left(\ph_0+\frac{\ph_+}{2}+\frac{\vph_-}{2}\right)+\cosh\left(\vph_0+\frac{\vph_+}{2}+\frac{\ph_-}{2}\right)\Big)f_1\non\\
&-&\sqrt{2m\s_2}\pp_{\ph_0}u_2^{-}(\p_++\chi_+)
\er
\br
\pp_+\ph_-&=&-\frac{m}{2}\left[ e^{\(\frac{\phi_+}{2}+\phi_0\)}\left(\s_1e^{\frac{\phi_-}{2}}+\s_2e^{-\frac{\phi_-}{2}}\) - e^{-\(\frac{\phi_+}{2}+\phi_0\)}\left(\s_1e^{-\frac{\phi_-}{2}}+\s_2e^{\frac{\phi_-}{2}}\)\right]\non\\&-&
\frac{i}{2}\sqrt{\frac{m}{2}}\left[(\sqrt{\s_1}u_1^{+}f_1+\sqrt{\s_2}u_2^{+}f_2)(\p_+-\chi_+)
+(\sqrt{\s_1}u_1^{-}g_1+\sqrt{\s_2}u_2^{-}g_2)(\p_++\chi_+)\right]\non\\
&-& \frac{im}{2}\sqrt{\s_1\s_2}\left[\sinh\left(\ph_0+\frac{\ph_+}{2}+\frac{\vph_-}{2}\right)f_1g_2+\sinh\left(\ph_0+\frac{\ph_+}{2}-\frac{\vph_-}{2}\right)g_1f_2\right]\\
\pp_-\ph_-&=&-\frac{m}{2}\left[e^{\(\frac{\phi_+}{2}-\phi_0\)}\left(\frac{1}{\s_1} e^{\frac{\phi_-}{2}}+\frac{1}{\s_2}e^{-\frac{\phi_-}{2}}\) - e^{-\(\frac{\phi_+}{2}-\phi_0\)}\left(\frac{1}{\s_1}e^{-\frac{\phi_-}{2}}+\frac{1}{\s_2}e^{\frac{\phi_-}{2}}\)\right]\non\\&+&
\frac{i}{2}\sqrt{\frac{m}{2}}\left[\(\frac{1}{\sqrt{\s_1}}v_1^{+}g_1-\frac{1}{\sqrt{\s_2}}v_2^{+}g_2\)(\bp_+-\bchi_+)
+\(\frac{1}{\sqrt{\s_1}}v_1^{-}f_1-\frac{1}{\sqrt{\s_2}}v_2^{-}f_2\)(\bp_++\bchi_+)\right]\non\\
&+& 
\frac{im}{2\sqrt{\s_1\s_2}}\left[\sinh\left(\ph_0-\frac{\ph_+}{2}+\frac{\vph_-}{2}\right)f_1g_2+\sinh\left(\ph_0-\frac{\ph_+}{2}-\frac{\vph_-}{2}\right)g_1f_2\right]\\
\pp_-(\ph_++2\ph_0)&=&-\frac{m}{2}\left[\,e^{\(\frac{\phi_+}{2}-\phi_0\)}\left(\frac{1}{\s_1} e^{\frac{\phi_-}{2}}-\frac{1}{\s_2}e^{-\frac{\phi_-}{2}}\) - e^{-\(\frac{\phi_+}{2}-\phi_0\)}\left(\frac{1}{\s_1}e^{-\frac{\phi_-}{2}}-\frac{1}{\s_2}e^{\frac{\phi_-}{2}}\)\right]\non\\
&+&
% % % %
\frac{i}{2}\sqrt{\frac{m}{2}}\left[\(\frac{1}{\sqrt{\s_1}}v_1^{+}g_1+\frac{1}{\sqrt{\s_2}}v_2^{+}g_2\)(\bp_+-\bchi_+)
+\(\frac{1}{\sqrt{\s_1}}v_1^{-}f_1+\frac{1}{\sqrt{\s_2}}v_2^{-}f_2\)(\bp_++\bchi_+)\right]\non\\
&-&
\frac{im}{2\sqrt{\s_1\s_2}}\left[\sinh\left(\vph_0-\frac{\vph_+}{2}+\frac{\ph_-}{2}\right)f_1g_2-\sinh\left(\vph_0-\frac{\vph_+}{2}-\frac{\ph_-}{2}\right)g_1f_2\right]\\
\pp_+(\ph_+-2\ph_0)&=&-\frac{m}{2}\left[ e^{\(\frac{\phi_+}{2}+\phi_0\)}\left(\s_1e^{\frac{\phi_-}{2}}-\s_2e^{-\frac{\phi_-}{2}}\) - e^{-\(\frac{\phi_+}{2}+\phi_0\)}\left(\s_1e^{-\frac{\phi_-}{2}}-\s_2e^{\frac{\phi_-}{2}}\)\right]\non\\&-&
% %
%
\frac{i}{2}\sqrt{\frac{m}{2}}\left[(\sqrt{\s_1}u_1^{+}f_1-\sqrt{\s_2}u_2^{+}f_2)(\p_+-\chi_+)
+(\sqrt{\s_1}u_1^{-}g_1-\sqrt{\s_2}u_2^{-}g_2)(\p_++\chi_+)\right]\non\\
&-& \frac{im}{2}\sqrt{\s_1\s_2}\left[\sinh\left(\vph_0+\frac{\vph_+}{2}+\frac{\ph_-}{2}\right)f_1g_2-\sinh\left(\vph_0+\frac{\vph_+}{2}-\frac{\ph_-}{2}\right)g_1f_2\right]\\
\pp_+\vph_-&=&-\frac{m}{2}\left[e^{\(\frac{\vph_+}{2}+\vph_0\)}\left(\s_1e^{\frac{\vph_-}{2}}+\s_2e^{-\frac{\vph_-}{2}}\) - e^{-\(\frac{\vph_+}{2}+\vph_0\)}\left(\s_1e^{-\frac{\vph_-}{2}}+\s_2e^{\frac{\vph_-}{2}}\)\right]\non\\&+&
\frac{i}{2}\sqrt{\frac{m}{2}}\left[(\sqrt{\s_1}u_1^{+}f_1+\sqrt{\s_2}u_2^{+}f_2)(\p_+-\chi_+)
-(\sqrt{\s_1}u_1^{-}g_1+\sqrt{\s_2}u_2^{-}g_2)(\p_++\chi_+)\right]\non\\
&+& \frac{im}{2}\sqrt{\s_1\s_2}\left[\sinh\left(\vph_0+\frac{\vph_+}{2}+\frac{\ph_-}{2}\right)f_1g_2+\sinh\left(\vph_0+\frac{\vph_+}{2}-\frac{\ph_-}{2}\right)g_1f_2\right]\\
\pp_-\vph_-&=&-\frac{m}{2}\left[e^{\(\frac{\vph_+}{2}-\vph_0\)}\left(\frac{1}{\s_1} e^{\frac{\vph_-}{2}}+\frac{1}{\s_2}e^{-\frac{\vph_-}{2}}\) - e^{-\(\frac{\vph_+}{2}-\vph_0\)}\left(\frac{1}{\s_1}e^{-\frac{\vph_-}{2}}+\frac{1}{\s_2}e^{\frac{\vph_-}{2}}\) \right]\non\\&-&
\frac{i}{2}\sqrt{\frac{m}{2}}\left[\(\frac{1}{\sqrt{\s_1}}v_1^{+}g_1-\frac{1}{\sqrt{\s_2}}v_2^{+}g_2\)(\bp_+-\bchi_+)
-\(\frac{1}{\sqrt{\s_1}}v_1^{-}f_1-\frac{1}{\sqrt{\s_2}}v_2^{-}f_2\)(\bp_++\bchi_+)\right]\non\\
&-& 
\frac{im}{2\sqrt{\s_1\s_2}}\left[\sinh\left(\vph_0-\frac{\vph_+}{2}+\frac{\ph_-}{2}\right)f_1g_2+\sinh\left(\vph_0-\frac{\vph_+}{2}-\frac{\ph_-}{2}\right)g_1f_2\right]\\
\pp_-(\vph_++2\vph_0)&=&-\frac{m}{2}\left[e^{\(\frac{\vph_+}{2}-\vph_0\)}\left(\frac{1}{\s_1} e^{\frac{\vph_-}{2}}-\frac{1}{\s_2}e^{-\frac{\vph_-}{2}}\) - e^{-\(\frac{\vph_+}{2}-\vph_0\)}\left(\frac{1}{\s_1}e^{-\frac{\vph_-}{2}}-\frac{1}{\s_2}e^{\frac{\vph_-}{2}}\) \right]\non\\
&-&
\frac{i}{2}\sqrt{\frac{m}{2}}\left[\(\frac{1}{\sqrt{\s_1}}v_1^{+}g_1+\frac{1}{\sqrt{\s_2}}v_2^{+}g_2\)(\bp_+-\bchi_+)
-\(\frac{1}{\sqrt{\s_1}}v_1^{-}f_1+\frac{1}{\sqrt{\s_2}}v_2^{-}f_2\)(\bp_++\bchi_+)\right]\non\\
&+& 
\frac{im}{2\sqrt{\s_1\s_2}}\left[\sinh\left(\ph_0-\frac{\ph_+}{2}+\frac{\vph_-}{2}\right)f_1g_2-\sinh\left(\ph_0-\frac{\ph_+}{2}-\frac{\vph_-}{2}\right)g_1f_2\right]
\er
\br
\pp_+(\vph_+-2\vph_0)&=&-\frac{m}{2}\left[e^{\(\frac{\vph_+}{2}+\vph_0\)}\left(\s_1e^{\frac{\vph_-}{2}}-\s_2e^{-\frac{\vph_-}{2}}\) - e^{-\(\frac{\vph_+}{2}+\vph_0\)}\left(\s_1e^{-\frac{\vph_-}{2}}-\s_2e^{\frac{\vph_-}{2}}\)\right]\non\\&+&
% % %
% % %
\frac{i}{2}\sqrt{\frac{m}{2}}\left[(\sqrt{\s_1}u_1^{+}f_1-\sqrt{\s_2}u_2^{+}f_2)(\p_+-\chi_+)
-(\sqrt{\s_1}u_1^{-}g_1-\sqrt{\s_2}u_2^{-}g_2)(\p_++\chi_+)\right]\non\\&+&
\frac{im}{2}\sqrt{\s_1\s_2}\left[\sinh\left(\ph_0+\frac{\ph_+}{2}+\frac{\vph_-}{2}\right)f_1g_2-\sinh\left(\ph_0+\frac{\ph_+}{2}-\frac{\vph_-}{2}\right)g_1f_2\right]
\er
It was verified that these B\"acklund transformations correspond to the equations (\ref{mov geral 1}) for each fields.

%%%%%%%%%%%%%%%%%%%%%%%%%%%%%%%%%%%%%%%%%%%%%%%%%%%%%%%%%%%%%%%

\end{document}